\documentclass[12pt]{article}
\usepackage{setspace}
\usepackage{longtable}
\usepackage{graphicx}

\usepackage{fancyvrb}
\makeatletter
\newcommand{\m}[1]{%
  \begingroup
  \def\FV@Space{ }
  \mathcode`\_="8000 
  \do@@us 
  $#1$%
  \endgroup
}
\newcommand{\do@@us}{%
  \begingroup\lccode`~=`\_ \lowercase{\endgroup\let~}\sb
}
\makeatother

\usepackage{spverbatim}

\usepackage{lscape}
\usepackage{rotating}
\usepackage{tcolorbox}

\usepackage[section]{placeins}
\usepackage[top=1in, left=1in, right=1in, bottom=1in]{geometry}
\usepackage[parfill]{parskip}	
\usepackage{graphicx}
\usepackage{caption}
\usepackage{subcaption}
\usepackage[export]{adjustbox}
\usepackage{amssymb}
\usepackage{epstopdf}
\usepackage{bbm}
\usepackage{bm}
\usepackage{amssymb}
\usepackage{amsmath}
\usepackage{amsfonts}

\usepackage{lscape}
\usepackage{enumitem}
\usepackage{rotating}
\usepackage{setspace}
\usepackage{threeparttable}
\usepackage{booktabs}
\usepackage[compact]{titlesec}
\usepackage{lmodern}

\fontfamily{lmtt}\selectfont
\usepackage[T1]{fontenc}

\usepackage[nocitation]{apacite}
\usepackage{natbib}
\bibpunct{(}{)}{;}{a}{}{,}
\usepackage{hyperref}
\usepackage{titling}
\usepackage{pifont}
\usepackage[capposition=top]{floatrow}
\newcommand{\subtitle}[1]{%
  \posttitle{%
    \par\end{center}
    \begin{center}\large#1\end{center}
    \vskip0.5em}%
}
\usepackage{amsthm}
\usepackage{amsmath} 
\usepackage{multirow}

\usepackage{multicol}
\usepackage{mathrsfs}

  {\end{smallmatrix}\right)}

\def\ci{\perp\!\!\!\perp}

\defcitealias{observational2021observational}{Observational Studies 2021}

\usepackage{mathrsfs}

\usepackage{color}

\usepackage{array}
\newcolumntype{C}[1]{>{\centering\arraybackslash}m{#1}}
\begin{document}
\pagestyle{plain}

\newtheoremstyle{mystyle}
{\topsep}
{\topsep}
{\it}
{}
{\bf}
{.}
{.5em}
{}
\theoremstyle{mystyle}
\newtheorem{assumptionex}{Assumption}
\newenvironment{assumption}
  {\pushQED{\qed}\renewcommand{\qedsymbol}{}\assumptionex}
  {\popQED\endassumptionex}
\newtheorem{assumptionexp}{Assumption}
\newenvironment{assumptionp}
  {\pushQED{\qed}\renewcommand{\qedsymbol}{}\assumptionexp}
  {\popQED\endassumptionexp}
\renewcommand{\theassumptionexp}{\arabic{assumptionexp}$'$}

\newtheorem{assumptionexpp}{Assumption}
\newenvironment{assumptionpp}
  {\pushQED{\qed}\renewcommand{\qedsymbol}{}\assumptionexpp}
  {\popQED\endassumptionexpp}
\renewcommand{\theassumptionexpp}{\arabic{assumptionexpp}$''$}

\newtheorem{assumptionexppp}{Assumption}
\newenvironment{assumptionppp}
  {\pushQED{\qed}\renewcommand{\qedsymbol}{}\assumptionexppp}
  {\popQED\endassumptionexppp}
\renewcommand{\theassumptionexppp}{\arabic{assumptionexppp}$'''$}

\renewcommand{\arraystretch}{1.3}

\newcommand{\argmin}{\mathop{\mathrm{argmin}}}
\makeatletter
\newcommand{\grande}{\bBigg@{2.25}}
\newcommand{\enorme}{\bBigg@{5}}

\newcommand{\blind}{0}

\newcommand{\tit}{\Large \texttt{lmw}: Linear Model Weights for Causal Inference}

\if0\blind

{\title{\tit\thanks{We thank Bijan Niknam for helpful commments. This work was supported by the Alfred P. Sloan Foundation (G-2020-13946) and the Patient-Centered
Outcomes Research Institute (PCORI, ME-2022C1-25648).}}
\author{Ambarish Chattopadhyay\thanks{Stanford Data Science, Stanford University, 450 Jane Stanford Way Wallenberg, Stanford, CA 94305; email: \url{hsirabma@stanford.edu}.} \and Noah Greifer\thanks{Institute for Quantitative Social Sciences (IQSS), Harvard University, CGIS Knafel Building, Office K303, 1737 Cambridge Street, Cambridge, MA 02138; emai: \url{ngreifer@iq.harvard.edu.}}
\and Jos\'{e} R. Zubizarreta\thanks{Departments of Health Care Policy, Biostatistics, and Statistics, Harvard University, 180 A Longwood Avenue, Office 307-D, Boston, MA 02115; email: \url{zubizarreta@hcp.med.harvard.edu}.}
}

\date{} 

\maketitle
}\fi

\if1\blind
\title{\bf \tit}
\date{} 
\maketitle
\fi

\begin{abstract}
The linear regression model is widely used in the biomedical and social sciences as well as in policy and business research to adjust for covariates and estimate the average effects of treatments. 
Behind every causal inference endeavor there is \textcolor{black}{a hypothetical} randomized experiment.
However, in routine regression analyses in observational studies, it is unclear how well the adjustments made by regression approximate key features of \textcolor{black}{randomized} experiments, such as covariate balance, study representativeness, sample boundedness, and unweighted sampling. 
In this paper, we provide software to empirically address this question.
\textcolor{black}{We introduce the \texttt{lmw} package for \texttt{R} to compute the implied linear model weights and perform diagnostics for their evaluation.}
The weights are obtained as part of the design stage of the study; that is, without using outcome information.
The implementation is general and applicable, for instance, in settings with instrumental variables and multi-valued treatments; in essence, in any situation where the linear model is the vehicle for adjustment and estimation of average treatment effects with discrete-valued interventions.
\end{abstract}


\begin{center}
\noindent Keywords:
{Causal Inference; Linear Models; \texttt{R} Programming Language; Regression Modeling; Observational Studies}
\end{center}
\clearpage
\doublespacing

\singlespacing
\pagebreak
\tableofcontents
\doublespacing

\pagebreak
\section{Introduction}

In observational studies, the linear regression model is widely used to estimate the causal effect of a treatment, intervention, or exposure on an outcome variable.
Examples are ubiquitous across the biomedical and social sciences, as well as in policy and business research. 
\textcolor{black}{In common practice, the outcome variable is linearly explained by a treatment indicator and a series of control variables or covariates, and the model is fitted by least squares.}
The estimated coefficient is interpreted as the average effect of the treatment on the response variable, holding the other \textcolor{black}{covariates} constant (e.g., \citealt{angrist2008mostly}, Chapter 3; \citealt{imbens2015causal}, Section 12.2.4).

The linear regression model has been extensively studied.
Among its properties is that the least squares estimator is the best (minimum variance) linear unbiased estimator of the treatment indicator coefficient, provided the model is correctly specified. 
Furthermore, the model is easy to fit and implement in practice, and is broadly accepted across disciplines as a valid method for data analysis.
However, when it comes to estimating the causal effect of a treatment, typical use of regression can blur certain notions and principles that are central to causal inference in observational studies.

One of them is the idea of approximating or emulating the randomized experiment that would have been conducted under controlled circumstances \citep{dorn1953philosophy, cochran1965planning, rosenbaum2010design1, hernan2016using}. 
\textcolor{black}{A key feature of randomized experimentation is the construction of treatment and control groups that are comparable or balanced in terms of confounding factors.}
Also central is a clear definition of the target population to which the inferences apply.
\textcolor{black}{Further, is the ideal of a study that ``self-analyzes;'' that is, where treatment effects can easily be investigated through simple, unweighted analyses.}
Finally, another important consideration is the extent to which the estimator extrapolates beyond the region \textcolor{black}{demarcated} by the observed data\textcolor{black}{, relying} on parametric assumptions that are untestable. 
In routine practice, it is unclear how linear regression adheres to these \textcolor{black}{standards}.

Arguably, other methods for adjustment in observational studies, such as matching and weighting, \textcolor{black}{align with these standards} more explicitly than regression in common practice.
For instance, weighting adjusts the treatment and control groups in order to balance their observed covariate distributions, characterizing the target population in an open manner, and producing an estimator that is sample bounded.
Moreover, weighting does not require outcome information to complete the adjustment process.
As argued by \cite{rubin2007design}, this helps to preserve the objectivity and validity of an observational study.

Recently, \cite{chattopadhyay2023implied} offered a connection between weighting and linear regression methods for causal inference.
In particular, they obtained new closed\textcolor{black}{-}form expressions for the implicit weights in different methods involving the linear regression model.
See \cite{abadie2015comparative}, \cite{imbens2015matching}, \cite{gelman2018high}, and \cite{ben2021augmented} for related analyses.
\cite{chattopadhyay2023implied} examined \textcolor{black}{the} properties of the implied weights of linear regression in both finite- and large-sample regimes.
In finite samples, they showed that the implied weights of general regression problems can be obtained equivalently by solving a convex optimization problem \textcolor{black}{similar} to the \textcolor{black}{one solved by the} stable balancing weights \citep{zubizarreta2015stable}, which involve minimizing the variance of the weights subject to constraints on covariate balance. 
In large samples, they studied multiply robust properties of regression estimators from the perspective of their implied weights. 
Among others, they proposed new regression diagnostics for causal inference that are part of the design stage of an observational study.

In this paper, we implement and demonstrate this weighting approach to regression.
In particular, we illustrate the new \emph{linear model weights} \texttt{lmw} package for \texttt{R}, which is freely available on CRAN.
Using it, we explore a number of settings.
For this, in Section 2 we present the notation, estimands, and basic assumptions.
In Section 3, we discuss two regression approaches, termed uni- and multi-regression imputation.
In Section 4, we introduce our running example.
In Section 5, we illustrate the use of the \texttt{lmw} package in settings with binary treatments, multivariate matching, multi-valued treatments, and instrumental variables, explaining step by step the code.
\textcolor{black}{In Section 6, we discuss associated \texttt{R} packages.}




\section{Setup}
\label{sec_setup}

Consider an observational study \textcolor{black}{with} $n$ independent and identically distributed units \textcolor{black}{randomly} drawn from a population.
For each unit $i \in \{1,2,...,n\}$, let $A_i$ be their treatment assignment indicator, with $A_i = 1$ if the unit is assigned to treatment, and $A_i = 0$ otherwise; $\bm{X}_i$ be a vector of $k$ observed covariates; and $Y_i$ be the observed outcome.
Define $n_t$ and $n_c$ as the sample sizes of the treatment and control groups, respectively.
Denote $\bar{\bm{X}}_t$ and $\bm{V}_t$, and $\bar{\bm{X}}_c$ and $\bm{V}_c$, as the mean vector and covariance matrix of the covariates in the treatment and control groups, correspondingly.
Write $\bar{\bm{X}}$ for the mean vector of the covariates in the entire sample.

Following the potential outcome framework (\citealt{neyman1923application, rubin1974estimating}), let $Y_i(a)$ be the potential outcome of unit $i$ under treatment $a \in \{0,1\}$ and $Y^{\text{obs}}_i = A_i Y_i(1) + (1-A_i)Y_i(0)$ be their observed outcome. 
In this paper, we are interested in two causal estimands: the average treatment effect \textcolor{black}{on the overall population}, $\text{ATE} = \mathbb{E}[Y_i(1) - Y_i(0)]$, and the average treatment effect on the treated, $\text{ATT} = \mathbb{E}[Y_i(1) - Y_i(0)|A_i=1]$. 
\textcolor{black}{To identify these estimands,} for the most of this paper we assume the following conditions hold \citep{rosenbaum1983central}: (a) positivity: $0<\Pr(A_i = 1\mid \bm{X}_i = \bm{x})<1$ for all $\bm{x} \in \text{supp}(\bm{X}_i)$, and (b) unconfoundedness: $A_i \ci \{Y_i(0),Y_i(1)\} \mid \bm{X}_i$.

\textcolor{black}{The first assumption, also known as the ``common support'' or ``overlap'' assumption, says that every unit, given their observed covariates, has a non-zero probability of being assigned to either the treatment or control group.
This assumption is crucial for avoiding extrapolation beyond the observed data and ensuring that effect estimates are grounded on the data as opposed to depending on further modeling assumptions.
The second assumption, also referred to as the ``ignorability'' or ``exchangeability'' assumption, posits that treatment assignment is independent of the potential outcomes given the observed covariates.
This assumption implies that, conditional on the observed covariates, any differences in outcomes between treatment groups can be attributed to the treatment.
This assumption motivates achieving balance in observed covariate distributions across treatment groups.
However, diagnostics for covariate balance reveal only that the observed variables are similarly distributed across groups, but do not account for unobserved covariates that may also affect the outcome.
Both assumptions are critical for drawing valid causal inferences in most of this paper, although in Sections 3.3 and 5.5 we discuss estimation with instrumental variables.
}

\section{Methods}
\label{sec_methods}

Following \cite{chattopadhyay2023implied}, we focus on two common regression modeling approaches. 
In the first approach, the outcome is regressed on the treatment indicator and the covariates without any treatment-covariate interactions, and the average treatment effect is estimated by the coefficient of the treatment indicator. 
This approach is called uni-regression imputation (URI) because it is equivalent to an imputation estimator based on a single (uni) linear outcome model. 
In the second approach, the outcome is regressed on a similar model that \textcolor{black}{includes} treatment-covariate interactions. 
This approach is called multi-regression imputation (MRI) because it is equivalent to an imputation estimator based on two (multi) linear outcome models, one fitted in each treatment group.

\subsection{Uni-regression imputation}
\label{sec_methods_URI}
URI is arguably the most commonly used method for regression adjustments in observational studies (\citealt{aronow2016does}, \citealt{angrist2008mostly} Chapter 3). 
The URI estimator of the ATE is \textcolor{black}{equivalent to} the ordinary least squares (OLS) estimator \textcolor{black}{$\hat{\tau}^{\scriptscriptstyle\text{OLS}}$} of $\tau$ in the model $Y^{\text{obs}}_{i} = \beta_0 + \bm{\beta}^\top_1 \bm{X}_i + \tau A_i + \epsilon_i$.\footnote{\textcolor{black}{More generally, we can consider URI models that regress the outcome on the treatment variable and arbitrary transformations of the covariate vector, e.g., higher-order interactions, basis functions corresponding to a kernel. That is, we can fit the model $Y^{\text{obs}}_{i} = \beta_0 + \tilde{\bm{\beta}}^\top_1 \bm{B}(\bm{X}_i) + \tau A_i + \epsilon_i$, where $\bm{B}(\bm{x}) = (B_1(\bm{x}),...,B_K(\bm{x}))^\top$ are $K$ arbitrary basis functions.}} 
Implementations of URI and associated inferences are straightforward and can be done using, e.g., the \texttt{lm} function in \texttt{R}. 
However, standard software implementations do not make clear how URI \textcolor{black}{uses the individual-level data to produce an aggregate effect estimate and how well it approximates certain features of randomized experiments in observational study designs}. 

To address this \textcolor{black}{challenge, we turn to} the implied weighting framework of \cite{chattopadhyay2023implied}. 
In this framework, the URI estimator is represented as a weighting estimator, $\hat{\tau}^{{\scriptscriptstyle \text{OLS}}}=  \sum_{i:A_i=1}w^{{\scriptscriptstyle \text{URI}}}_i Y^{\text{obs}}_i - \sum_{i:A_i=0}w^{{\scriptscriptstyle \text{URI}}}_i Y^{\text{obs}}_i$, where $w^{{\scriptscriptstyle \text{URI}}}_i = \frac{1}{n_t} + (\bm{X}_i  -\bar{\bm{X}}_t)^\top (n_t\bm{V}_t + n_c\bm{V}_c)^{-1} (\bar{\bm{X}}_c - \bar{\bm{X}}_t)$ if unit $i$ is in the treatment group and $w^{{\scriptscriptstyle \text{URI}}}_i = \frac{1}{n_c} + (\bm{X}_i  -\bar{\bm{X}}_c)^\top (n_t\bm{V}_t + n_c\bm{V}_c)^{-1} (\bar{\bm{X}}_t - \bar{\bm{X}}_c)$ if unit $i$ is in the control group. 
The weights $w^{{\scriptscriptstyle \text{URI}}}_i$ sum to one within each group and are functions of the covariates but not the outcome.\footnote{\textcolor{black}{Hence, the weights are agnostic to the type of outcome, discrete, continuous, or bounded.}} 
These weights make precise how regression acts on treated and control units in the sample without using outcome information. 
As a result, we can characterize in finite samples the adjustments \textcolor{black}{made by} URI in terms of covariate balance, study representativeness, model extrapolation, and weight dispersion. 

In particular, the URI weights exactly balance the means of the covariates, but towards \textcolor{black}{an implied} population that is different from the ``natural'' populations corresponding to the treated population, the control population, or the overall population.
\textcolor{black}{The population that URI targets is characterized by the covariate profile $\bm{X}^{*{\scriptscriptstyle \text{URI}}} := n_c\bm{V}_c(n_t\bm{V}_t + n_c\bm{V}_c)^{-1} \bar{\bm{X}}_t + n_t\bm{V}_t (n_t\bm{V}_t + n_c\bm{V}_c)^{-1} \bar{\bm{X}}_c$ where $\bm{X}^{*{\scriptscriptstyle \text{URI}}} = \sum_{i:A_i=1} w^{{\scriptscriptstyle \text{URI}}}_i \bm{X}_i = \sum_{i:A_i=0} w^{{\scriptscriptstyle \text{URI}}}_i \bm{X}_i$.
This reveals a potential limitation of URI, because if the model is misspecified and the treatment effect is, in fact, modified by covariates, then the effect estimator will be biased.
In comparison to other methods, with matching and weighting the population after covariate adjustments is often explicitly characterized; however, if there is lack of overlap in covariate distributions across treatment groups, this population may not align with the treated population or the other natural populations.\footnote{See, for example, \cite{cohn2021profile} and \cite{chattopadhyay2023one} for matching and weighting methods that target populations of specific interest.} 
Consequently, we recommend the use of target standardized mean differences (TSMDs) to assess the balance of the groups, not only amongst themselves, but also in relation to the target population \citep{chattopadhyay2020balancing}.}

\textcolor{black}{Another relevant dimension is that the URI weights can be negative, which may lead to extrapolation beyond the support of the observed data. In other words, with negative weights, weighted estimators of the form $\sum_{i:A_i = a}w_i Y^{\text{obs}}_i$ are not guaranteed to lie within the range of the observed outcome data, i.e., $[\min\{Y^{\text{obs}}_i: A_i =a\},\max\{ Y^{\text{obs}}_i: A_i =a\}],$ and thus, they are not sample-bounded in the sense of \cite{robins2007comment}.}
Finally, the URI weights are optimal in that they are the weights of minimum variance that sum to one and exactly balance the means of the covariates towards the profile $\bm{X}^{*{\scriptscriptstyle \text{URI}}}$. 


\subsection{Multi-regression imputation}
\label{sec_methods_MRI}

MRI is another widely used method for regression adjustments in observational studies, also known as g-computation or regression estimation in the causal inference literature (\citealt{hernan2020causal}, \citealt{schaferAverageCausalEffects2008}). 
In MRI, we fit a linear regression model $Y^{\text{obs}}_i = \beta_{0t} + \bm{\beta}^\top_{1t} \bm{X}_i + \epsilon_{it}$ in the treatment group and a separate linear regression model $Y^{\text{obs}}_i = \beta_{0c} + \bm{\beta}^\top_{1c} \bm{X}_i + \epsilon_{ic}$ in the control group using OLS.\footnote{\textcolor{black}{More generally, we can consider MRI models that regress the outcome on arbitrary transformations of the covariate vector. For instance, we can fit the model $Y^{\text{obs}}_{i} = \beta_{0t} + \tilde{\bm{\beta}}^\top_{1t} \bm{B}(\bm{X}_i) + \epsilon_{it}$ in the treatment group, and the model $Y^{\text{obs}}_{i} = \beta_{0c} + \tilde{\bm{\beta}}^\top_{1c} \bm{B}(\bm{X}_i) + \epsilon_{ic}$ in the control group, where $\bm{B}(\bm{x}) = (B_1(\bm{x}),...,B_K(\bm{x}))^\top$ are $K$ arbitrary basis functions.}} The MRI estimators of the ATE and the ATT are then $\widehat{\text{ATE}} = \frac{1}{n}\sum_{i=1}^{n} (\hat{\beta}_{0t} + \hat{\bm{\beta}}^\top_{1t} \bm{X}_i) - \frac{1}{n}\sum_{i=1}^{n} (\hat{\beta}_{0c} + \hat{\bm{\beta}}^\top_{1c} \bm{X}_i)$ and $\widehat{\text{ATT}} = \frac{1}{n}\sum_{i:A_i=1}^{n} Y^{\text{obs}}_i - \frac{1}{n_t}\sum_{i:A_i=1} (\hat{\beta}_{0c} + \hat{\bm{\beta}}^\top_{1c} \bm{X}_i)$.

\textcolor{black}{Using the implied weighting framework, we can represent the MRI estimators for the estimands of interest as weighting estimators, of the form $ \sum_{i:A_i=1}w^{{\scriptscriptstyle \text{MRI}}}_{i} Y^{\text{obs}}_i - \sum_{i:A_i=0}w^{{\scriptscriptstyle \text{MRI}}}_{i} Y^{\text{obs}}_i$, where the weights sum to one within each group (see Proposition 2 in \citealt{chattopadhyay2023implied}). The closed-form expressions of the weights depend on the target estimand.} 
For the ATE, $w^{{\scriptscriptstyle \text{MRI}}}_{i} = \frac{1}{n_t}\{1 + (\bm{X}_i -\bar{\bm{X}}_t)^\top \bm{V}_t^{-1} (\bar{\bm{X}} - \bar{\bm{X}}_t)\}$ if unit $i$ is treated and $w^{{\scriptscriptstyle \text{MRI}}}_{i} = \frac{1}{n_c}\{1 + (\bm{X}_i -\bar{\bm{X}}_c)^\top \bm{V}_c^{-1} (\bar{\bm{X}} - \bar{\bm{X}}_c)\}$ otherwise. 
For the ATT, $w^{{\scriptscriptstyle \text{MRI}}}_{i} = \frac{1}{n_t}$ if unit $i$ is treated and $w^{{\scriptscriptstyle \text{MRI}}}_{i} = \frac{1}{n_c}\{1 + (\bm{X}_i -\bar{\bm{X}}_c)^\top \bm{V}_c^{-1} (\bar{\bm{X}}_t - \bar{\bm{X}}_c)\}$ otherwise. Similar to URI, the MRI weights do not depend on the outcome, so they can also be obtained in the design stage of a causal study. Specifically, the MRI weights exactly balance the means of the covariates, but unlike the URI weights, they balance them toward the intended target populations. 
That is, $\sum_{i:A_i=1} w^{{\scriptscriptstyle \text{MRI}}}_i \bm{X}_i = \sum_{i:A_i=0} w^{{\scriptscriptstyle \text{MRI}}}_i \bm{X}_i = \bm{X}^{*{\scriptscriptstyle \text{MRI}}}$ where $\bm{X}^{*{\scriptscriptstyle \text{MRI}}}$ equals  $\bar{\bm{X}}$ and $\bar{\bm{X}}_t$ when the estimand is the ATE and the ATT, respectively. 
Further, the MRI weights the weights of minimum variance that sum to one and exactly balance the means of the covariates towards the profile $\bm{X}^{*{\scriptscriptstyle \text{MRI}}}$. 
Finally, similar to URI, the MRI weights can take negative values, leading to possible extrapolation beyond the support of the observed data. 


In large samples, both URI and MRI weights converge to inverse probability weights, albeit under stringent assumptions on the treatment assignment process. Moreover, leveraging this convergence, the URI and MRI estimators can be shown to be multiply robust under conditions for the true treatment and outcome models. See \cite{chattopadhyay2023implied} for details.

\subsection{Extensions \textcolor{black}{to multi-valued treatments and instrumental variables}}
\label{sec_methods_extensions}

The implied weighting framework applies to regression estimators in more general settings, such as those with multi-valued treatments and instrumental variables. 
First, consider a multi-valued treatment taking values in a finite set $\mathcal{A}$ of treatment levels. Without loss of generality, let the estimand be $\text{ATE}_{a,1} := \mathbb{E}[Y_i(a) - Y_i(1)]$, $a \in \{2,...,|\mathcal{A}|\}$, i.e., the average treatment effect of level $a$ (the active level) over level $1$ (the reference level). The MRI approach fits separate linear regression models in the groups $A_i = a$ and $A_i=1$ and hence, the MRI weights and their properties for binary treatment settings directly apply to this setting. The URI approach fits the model $Y^{\textrm{obs}}_i = \beta_0 + \bm{\beta}^\top_1 \bm{X}_i + \sum_{a=2}^{|\mathcal{A}|}\tau_{a,1} \mathbbm{1}(A_i=a) + \epsilon_i $ and estimates $\text{ATE}_{a,1}$ by $\hat{\tau}^{\scriptscriptstyle\text{OLS}}_{a,1}$. This estimator admits the weighting representation $\hat{\tau}^{\scriptscriptstyle \text{OLS}}_{a,1} = \sum_{i:A_i=a}w^{{\scriptscriptstyle \textrm{URI}}}_{i,a,1} Y^{\textrm{obs}}_i - \sum_{A_i \neq a} w^{{\scriptscriptstyle \textrm{URI}}}_{i,a,1} Y^{\textrm{obs}}_i$, where the weights $w^{{\scriptscriptstyle \textrm{URI}}}_{i,a,1}$ sum to one within the groups $A_i = a$ and $A_i \neq a$. In general, the URI weights may be non-zero for units in groups other than groups $a$ and $1$. 
That is, other treatment groups besides $a$ and $1$ have a say in the contrast between groups $a$ and $1$.
In fact, when estimating $\text{ATE}_{a,1}$, URI borrows strength from all the treatment groups using linearity. Finally, while the weights exactly balance the means of the covariates between groups $A_i = a$ and $A_i \neq a$, the implied target population under URI varies with the active treatment level $a$ (that is, it varies within the same fitted model), and in general, none of these populations match the overall study population. 

Second, consider an instrumental variable (IV) setting with a binary instrument $Z_i$, which takes the value $Z_i = 1$ if unit $i$ is encouraged to receive treatment, and $Z_i = 0$ otherwise. A common method to estimate the effect of treatment in this setting is two-stage least squares (2SLS) regression. In the first stage of 2SLS, one fits the regression model $A_i = \delta_0 + \bm{\delta}^\top_1 \bm{X}_i + \gamma Z_i + \epsilon_i$ using OLS. In the second stage, one uses $\hat{A}_i$, the predicted values of $A_i$ from the first stage, to fit the regression model $Y^{\textrm{obs}}_i = \alpha_0 + \bm{\alpha}^\top_1 \bm{X}_i + \tau \hat{A}_i + \eta_i$ again using OLS. The 2SLS estimator is $\hat{\tau}^{\scriptscriptstyle\text{2SLS}}$, the estimator of $\tau$ from the second stage regression. Using the implied weighting framework, $\hat{\tau}^{\scriptscriptstyle\text{2SLS}}$ can also be represented as a weighting estimator: 
$\hat{\tau}^{{\scriptscriptstyle \text{2SLS}}} = \sum_{i:A_i=1} w^{{\scriptscriptstyle \textrm{2SLS}}}_i Y^{\textrm{obs}}_i - \sum_{i:A_i=0} w^{{\scriptscriptstyle \textrm{2SLS}}}_i Y^{\textrm{obs}}_i$. 
\textcolor{black}{The vector of weights $\bm{w}^{{\scriptscriptstyle \text{2SLS}}} = (w^{{\scriptscriptstyle \text{2SLS}}}_1,...,w^{{\scriptscriptstyle \text{2SLS}}}_n)^\top$ can be written in closed form as $\bm{w}^{{\scriptscriptstyle \text{2SLS}}} =  (2\underline{\bm{A}} - \underline{\bm{I}})\frac{(\underline{\bm{I}} - \underline{\bm{\mathcal{P}_X}})\bm{Z}}{\bm{Z}^\top(\underline{\bm{I}} - \underline{\bm{\mathcal{P}_X}})\bm{A}}$, where $\bm{Z} = (Z_1,...,Z_n)^\top$, $\bm{A} = (A_1,...,A_n)^\top, \underline{\bm{A}} = \text{diag}(A_1,...,A_n)$, $\underline{\bm{I}}$ is the identity matrix, and $\underline{\bm{\mathcal{P}_X}}$ is the projection matrix onto the column space of the augmented covariate matrix (i.e., the matrix of covariates, augmented with a column of all 1s).}
\textcolor{black}{Thus, the weights depend only on the covariates, treatment, and the instrument.}
\textcolor{black}{More generally, a similar weighting representation holds for two-stage predictor substitution (2SPS) methods \citep{terza2008two}, when the outcome model is additive in the treatment variable and arbitrary basis functions of the covariates.}
\textcolor{black}{Similar} to the URI and MRI weights, the 2SLS weights \textcolor{black}{sum to one within each treatment group and} exactly balance the means of the covariates. Moreover, these weights allow us to explicitly characterize the implied target populations of 2SLS regression in finite samples. In general, the implied population of 2SLS differs from the treated, control, or the overall study population, which is why the 2SLS estimand, the local average treatment effect (LATE), often differs from the true ATE. Finally, the weights under 2SLS can be negative and, in fact, such negative weights occur more frequently in this setting than under URI (\citealt{chattopadhyay2022new, chattopadhyay2023implied}). 

\section{Data}
\label{sec_data}

To illustrate the \texttt{lmw} package, we use as a running example a part of the famous Lalonde observational study (\citealt{lalonde1986evaluating}). The study evaluates the effect of a labor training program 
(the National Supported Work Demonstration) on future earnings. In our analysis, the treatment group corresponds to a subset of $n_t = 185$ units enrolled in the program from the Lalonde experimental study \citep{dehejia1999causal}, and the control group corresponds to a subset of $n_c = 2490$ non-experimental units from the Population Survey of Income Dynamics (PSID). The dataset can be downloaded from \url{https://users.nber.org/~rdehejia/data/.nswdata2.html}.
There are $k=7$ baseline covariates: \texttt{age} (age in years), \texttt{education} (years of education), \texttt{race} (Black, Hispanic, and White), \texttt{married} (indicator for married status), \texttt{nodegree} (indicator for high school dropout), \texttt{re74} (earnings in 1974), and \texttt{re75} (earnings in 1975). The outcome variable of interest is \texttt{re78} (earnings in 1978). The standard estimand in this study is the ATT, i.e., the average effect of the training program on the earnings of subjects who are part of the program. However, for illustration, we also consider estimating the ATE.

In addition, we use the covariate data to generate an artificial multi-valued treatment with three levels. We label the original treatment group as group 1 and split the original control group into two groups: group 2 (of size 901) and group 3 (of size 1589) based on the logistic regression model $\text{logit}\{P(A_i = 3|\bm{X}_i, A_i \neq 1)\} = 1 - \texttt{age} + 1.2 \times  \texttt{education} - \texttt{black} + 1.2 \times \texttt{re75}$. We also generate an artificial binary instrument $Z_i$ based on the logistic regression model $\text{logit}\{P(Z_i = 1|\bm{X}_i, A_i \neq 1)\} = 1 - \texttt{age}  + 0.5 \times \texttt{education} - 0.5 \times \texttt{hispanic} + 0.8 \times \text{black} + 0.8 \times \texttt{re74} + 3 \times \mathbbm{1}(A_i=1)$. The generated instrument, by construction, has a significant partial correlation with treatment after adjusting for the covariates. Also, the instrument passes a pseudo-outcome-based heuristic test for the assumptions of exclusion restriction and independence with unobserved covariates (see \citealt{baiocchi2014instrumental}). 

\section{Using \texttt{lmw}}

Although the usual target estimand with the Lalonde data is the ATT, here we target the ATE to demonstrate the features of \texttt{lmw} that can be used to diagnose balance, representativeness, and extrapolation. We \textcolor{black}{then} follow with an example targeting the ATT, an example using a multi-valued treatment, and an example using 2SLS with an instrumental variable.

The primary function of the \texttt{lmw} package is \texttt{lmw()}, which computes the implied regression weights. The syntax of \texttt{lmw()} is very similar to the linear modeling function \texttt{lm()}, with a few adjustments.

\singlespacing
\begin{verbatim}
lmw_out <- lmw(~ A + X1 + X2 + X3, data = dataset,
               estimand = "ATE", method = "URI",
               treat = "A")
\end{verbatim}
\doublespacing

The first argument specifies the regression model formula, with the treatment and covariates on the right-hand-side of \texttt{\~}. Unlike with \texttt{lm()}, an outcome does not need to be specified because the linear regression weights do not depend on the outcome; this allows for the preservation of the distinction between the design and analysis stages.
The second argument, \texttt{data}, specifies the dataset containing the variables in the model formula. The \texttt{estimand} argument specifies the desired target estimand for the analysis. The \texttt{method} argument specifies whether URI or MRI is to be used. Finally, the \texttt{treat} argument specifies which variable in the model formula is to be regarded as the treatment variable; if omitted, the first variable is assumed to be the treatment. \texttt{lmw()} has several other arguments for more advanced uses that will be discussed later.

\subsection{URI for the ATE}
Below is an example of using \texttt{lmw()} to compute the URI weights for the Lalonde dataset. After running \texttt{lmw()}, the output is printed, displaying a summary of the resulting \texttt{lmw} object.

\singlespacing
\begin{verbatim}
lmw_ate_out <- lmw(~ treat + age + education + married + race + 
                     nodegree + re74 + re75, data = lalonde,
                   estimand = "ATE", method = "URI", treat = "treat")
lmw_ate_out

## An lmw object
##  - treatment: treat (2 levels)
##  - method: URI (uni-regression imputation)
##  - number of obs.: 2675
##  - sampling weights: none
##  - base weights: none
##  - target estimand: ATE
##  - covariates: age, education, married, race, nodegree, re74, re75
\end{verbatim}
\doublespacing
The next step is to run \texttt{summary()} on the resulting object to examine the balance and representativeness of the sample weighted by the \textcolor{black}{implied} regression weights.

\singlespacing
\begin{verbatim}
lmw_ate_sum <- summary(lmw_ate_out)
lmw_ate_sum
## 
## Call:
## lmw(formula = ~ treat + age + education + married + race + nodegree + 
##     re74 + re75, data = lalonde, estimand = "ATE", method = "URI", 
##     treat = "treat")
## 
## Summary of Balance for Unweighted Data:
##                 SMD TSMD Control TSMD Treated    KS TKS Control TKS Treated
## age          -1.009        0.070       -0.940 0.377       0.026       0.351
## education    -0.681        0.047       -0.633 0.403       0.028       0.375
## married      -1.845        0.128       -1.718 0.677       0.047       0.630
## raceblack     1.482       -0.102        1.379 0.593       0.041       0.552
## racehispanic  0.129       -0.009        0.120 0.027       0.002       0.025
## racewhite    -1.625        0.112       -1.512 0.620       0.043       0.577
## nodegree      0.880       -0.061        0.820 0.403       0.028       0.375
## re74         -1.718        0.119       -1.599 0.729       0.050       0.679
## re75         -1.774        0.123       -1.652 0.774       0.054       0.720
## 
## Summary of Balance for Weighted Data:
##              SMD TSMD Control TSMD Treated    KS TKS Control TKS Treated
## age            0       -0.891       -0.891 0.127       0.350       0.332
## education      0       -0.615       -0.615 0.081       0.352       0.352
## married        0       -1.574       -1.574 0.000       0.578       0.578
## raceblack      0        1.338        1.338 0.000       0.535       0.535
## racehispanic   0        0.129        0.129 0.000       0.027       0.027
## racewhite      0       -1.474       -1.474 0.000       0.562       0.562
## nodegree       0        0.769        0.769 0.000       0.352       0.352
## re74           0       -1.578       -1.578 0.578       0.528       0.665
## re75           0       -1.636       -1.636 0.566       0.523       0.711
## 
## Effective Sample Sizes:
##          Control Treated
## All       2490.    185. 
## Weighted   367.3   180.6
\end{verbatim}
\doublespacing
\textcolor{black}{The function} \texttt{summary()} produces two tables, one \textcolor{black}{with} balance statistics for the original unweighted sample and \textcolor{black}{another} one \textcolor{black}{with} balance statistics for the \textcolor{black}{weighted} sample by the regression weights. 
Each row \textcolor{black}{displays} a covariate, and each column \textcolor{black}{presents} a balance statistic. 
The statistics include \textcolor{black}{the commonly used} standardized mean difference (\texttt{SMD}) of \textcolor{black}{a} covariate between the treatment and control groups \textcolor{black}{(\citealt{rosenbaum1985constructing}, \citealt{austin2015moving}). 
Moreover, they include the target standardized mean difference} \texttt{TSMD Control} and \texttt{TSMD Treated}, \textcolor{black}{defined as the} SMD between each treatment group and the target population of the target estimand (in this case, the \textcolor{black}{overall} sample because \texttt{estimand} is set to \texttt{"ATE"}; see \citealt{chattopadhyay2020balancing}). 
\textcolor{black}{More formally, let $\bar{X}_1$ and $\bar{X}_{w,1}$ be the unweighted and weighted means of a covariate $X$ in the treated sample, respectively. Write $\bar{X}$ and $s_X$ as the mean and standard deviation of $X$ in the overall sample, respectively. When the estimand is the ATE, the TSMD of $X$ in the unweighted treated sample is $\frac{\bar{X}_1 - \bar{X}}{s_X}$, and in the weighted treated sample is $\frac{\bar{X}_{w,1} - \bar{X}}{s_X}$. The TSMD in the control sample is computed analogously. TSMD values closer to zero indicate better mean balance relative to the target. In practice, one may choose to report the absolute value of TSMD, also known as TASMD (target absolute standardized mean difference).}

\textcolor{black}{The balance statistics also include the Kolmogorov-Smirnov (KS) statistics between the treatment and control groups (\texttt{KS}; see, e.g., \citealt
{austin2015moving}), and between each group and the target (Target Kolmogorov-Smirnov or \texttt{TKS}). More formally, let $\hat{F}_{1}, \hat{F}_{w,1}$, and $\hat{F}$ be the empirical CDFs of a covariate $X$ in the unweighted treated sample, the weighted treated sample, and the overall sample, respectively. In this case, the TKS in the unweighted and weighted treated samples are $\sup\limits_{x \in \text{Support}(X)}|\hat{F}_1(x) - \hat{F}(x)|$ and $\sup\limits_{x \in \text{Support}(X)}|\hat{F}_{w,1}(x) - \hat{F}(x)|$, respectively. Smaller values of these statistics indicate better balance in the marginal distribution of $X$ relative to the target.} 
The \texttt{addlvariables} argument can be supplied with the names of additional variables or terms (e.g., transformations of the covariates or interactions between them) to assess balance on them in addition to the variables used in the model. 
\textcolor{black}{Finally,} the effective sample size (ESS) \textcolor{black}{is computed as $\left(\sum_{i:A_i = a} w_i\right)^2 / \sum_{i:A_i = a} w_i^2$ for each treatment group $a \in \{0, 1\}$} and represents the approximate size of an unweighted sample that contains the same precision as the weighted sample in question for estimating a counterfactual mean.

One can also use \texttt{plot()} on the \texttt{summary()} output to produce a visualization of the resulting balance. The plot can be customized to display difference balance statistics, add threshold lines, and change the order of the covariates.

\singlespacing
\begin{verbatim}
plot(lmw_ate_sum)
\end{verbatim}
\doublespacing

\begin{figure}[H]
    \centering
    \includegraphics[width=4.5in]{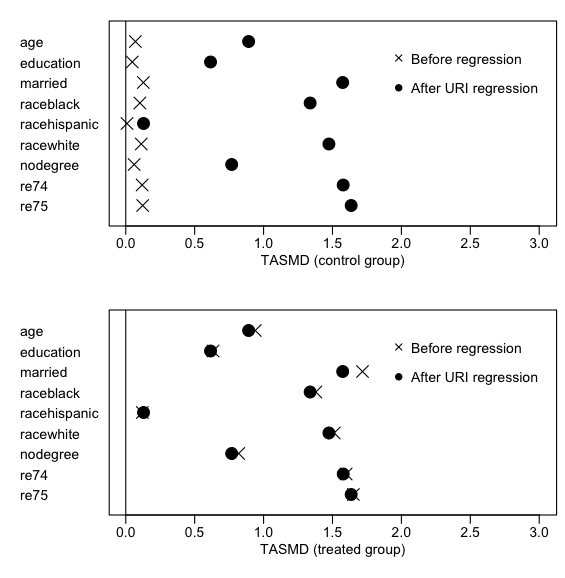}
\end{figure}

The first table in the \texttt{summary()} output reveals several imbalances between the treatment groups and between each treatment group and the target population before \textcolor{black}{regression}, as indicated by the high SMDs and KS statistics, especially for the treated group. 
\textcolor{black}{These imbalances} are also indicated by the crosses ($\times$) in the \textcolor{black}{above} plot. After regression, the SMDs are equal to \textcolor{black}{zero} for all covariates, indicating good covariate balance between the treatment groups (though some of the KS statistics remain high; \textcolor{black}{see the second table in the \texttt{summary()} output}). However, the target imbalance is severe, as indicated by the TSMD and TKS statistics \textcolor{black}{in the summary table} and the black circles \textcolor{black}{($\bullet$)} in the plots. In particular, target imbalance for the control group becomes much worse than it was \textcolor{black}{before regression}, while target imbalance for the treated group barely changes. 
The distribution of the covariates in the resulting weighted sample can be summarized using \texttt{summary()} with the additional argument \texttt{stat = "distribution"}:

\singlespacing
\begin{footnotesize}
\begin{verbatim}
summary(lmw_ate_out, stat = "distribution")

## Call:
## lmw(formula = ~ treat + age + education + married + race + nodegree + 
##     re74 + re75, data = lalonde, estimand = "ATE", method = "URI", 
##     treat = "treat")
## 
## Distribution Summary for Unweighted Data:
##              Mean Overall  SD Overall Mean Control  SD Control Mean Treated SD Treated
## age                34.226    (10.500)       34.851    (10.441)       25.816    (7.155)
## education          11.994     (3.054)       12.117     (3.082)       10.346    (2.011)
## married             0.819     (0.385)        0.866     (0.340)        0.189    (0.392)
## raceblack           0.292     (0.454)        0.251     (0.433)        0.843    (0.364)
## racehispanic        0.034     (0.182)        0.033     (0.177)        0.059    (0.236)
## racewhite           0.674     (0.469)        0.717     (0.451)        0.097    (0.296)
## nodegree            0.333     (0.471)        0.305     (0.461)        0.708    (0.455)
## re74            18230.003 (13722.252)    19428.746 (13406.877)     2095.574 (4886.620)
## re75            17850.894 (13877.777)    19063.338 (13596.955)     1532.055 (3219.251)
## 
## Distribution Summary for Weighted Data:
##              Mean Target   SD Target Mean Control  SD Control Mean Treated SD Treated
## age               34.226    (10.500)       26.247     (6.071)       26.247    (7.283)
## education         11.994     (3.054)       10.394     (2.821)       10.394    (2.053)
## married            0.819     (0.385)        0.242     (0.428)        0.242    (0.428)
## raceblack          0.292     (0.454)        0.827     (0.378)        0.827    (0.378)
## racehispanic       0.034     (0.182)        0.061     (0.240)        0.061    (0.240)
## racewhite          0.674     (0.469)        0.112     (0.315)        0.112    (0.315)
## nodegree           0.333     (0.471)        0.685     (0.465)        0.685    (0.465)
## re74           18230.003 (13722.252)     2310.339 (19516.079)     2310.339 (5243.971)
## re75           17850.894 (13877.777)     1690.432 (20103.272)     1690.432 (3525.011)
##
## Effective Sample Sizes:
##          Control Treated
## All       2490.    185. 
## Weighted   367.3   180.6
\end{verbatim}
\end{footnotesize}
\doublespacing

The first table contains the mean and standard deviation of each covariate in the \textcolor{black}{overall} sample and each treatment group prior to weighting, and the second table contains the mean and standard deviation of each covariate in the target population and in the treatment groups after weighting. The severe target imbalances are \textcolor{black}{also} apparent \textcolor{black}{from} this table. For example, the mean age in the target population is approximately 34 years, while in the sample weighted by the regression weights, the mean age is approximately 26 years. The proportion of married subjects in the target population is approximately 82\%, and in the weighted sample is 24\%. These extreme imbalances demonstrate that the population implied by the URI weights is not representative of the population that the ATE intends to target.

Using \texttt{plot()} directly on the \texttt{lmw} output object provides additional insight into the distribution of the weights. The \texttt{type} argument can be used to select which among \textcolor{black}{the} three types of plots to display; we consider the extrapolation plot (\texttt{type = "extrapolation"}) here and the other two later.

\singlespacing
\begin{verbatim}
plot(lmw_ate_out, type = "extrapolation",
     variables = ~ age + married + re74)
\end{verbatim}
\doublespacing

\begin{figure}[H]
    \centering
    \includegraphics[width=4.5in]{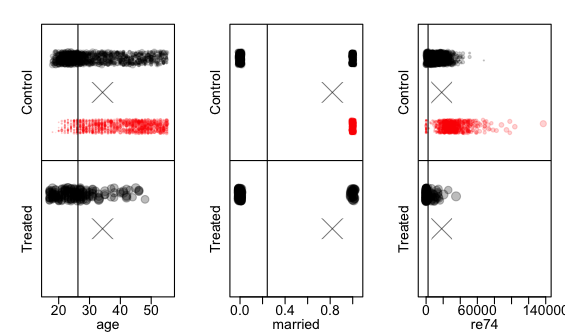}
\end{figure}

The extrapolation plot displays how the weights distribute by the covariates listed in \texttt{variables}. The vertical line indicates the mean of the covariate in each treatment group after weighting, and the cross ($\times$) indicates the mean of the covariate in the target population. Each point represents an observation, with the size of each point reflecting the magnitude of its weight. \textcolor{black}{To improve clarity, within each box, points are jittered vertically, with positively weighted (black) and negatively weighted (red) points separated; otherwise the vertical axis carries no information.} From this plot it is apparent that the covariate means in the treatment groups do not align with the means in the target population, and extrapolation is fairly severe given the preponderance of negative weights (red points). Negative weights often yield poor balance on components of the covariate distribution not directly targeted by the model; this can be seen in the high KS statistics in the weighted sample even while the SMDs are 0.

Given the severe target imbalance and extrapolation induced by using URI to estimate the ATE, one has a few options. First, one can attempt to use other methods that retain the target estimand, such as MRI or other weighting or matching methods (possibly in combination with MRI or URI). Second, one can change the target estimand of interest to one that is better supported by the data, though this choice should be determined primarily by substantive considerations (see \citealt{greifer2021choosing}). We use both of these approaches below, changing the target estimand to the study’s more natural estimand, the ATT, and investigating the performance of MRI and an initial matching step to reduce extrapolation and dependence on the outcome model.

\subsection{MRI for the ATT}
\label{sec_mri_att}
Below, we use the MRI to target the ATT. The call to \texttt{lmw()} is almost identical, except this time we set \texttt{method = "MRI"} and  \texttt{estimand = "ATT"}:

\singlespacing
\begin{verbatim}
lmw_att_out <- lmw(~ treat + age + education + married + race + 
                     nodegree + re74 + re75, data = lalonde,
                     estimand = "ATT", method = "MRI", treat = "treat")
lmw_att_out
## An lmw object
##  - treatment: treat (2 levels)
##  - method:  MRI (multi-regression imputation)
##  - number of obs.: 2675
##  - sampling weights: none
##  - base weights: none
##  - target estimand: ATT
##  - covariates: age, education, married, race, nodegree, re74, re75
\end{verbatim}
\doublespacing

The model formula is interpreted slightly differently when using MRI than when using URI. With URI, the model formula corresponds exactly to the linear model that a user would fit, e.g., using \texttt{lm()}; with MRI, the model formula simply indicates the terms that will be fit in the linear model separately for each treatment group, and the treatment variable can be omitted (but still must be specified in the treat argument).

After computing the weights, target balance can be assessed using \texttt{summary()}:

\singlespacing
\begin{verbatim}
lmw_att_sum <- summary(lmw_att_out)
lmw_att_sum
## 
## Call:
## lmw(formula = ~treat + age + education + married + race + nodegree + 
##     re74 + re75, data = lalonde, estimand = "ATT", method = "MRI", 
##     treat = "treat")
## 
## Summary of Balance for Unweighted Data:
##                 SMD TSMD Control TSMD Treated    KS TKS Control TKS Treated
## age          -1.263        1.263            0 0.377       0.377           0
## education    -0.881        0.881            0 0.403       0.403           0
## married      -1.729        1.729            0 0.677       0.677           0
## raceblack     1.630       -1.630            0 0.593       0.593           0
## racehispanic  0.114       -0.114            0 0.027       0.027           0
## racewhite    -2.091        2.091            0 0.620       0.620           0
## nodegree      0.886       -0.886            0 0.403       0.403           0
## re74         -3.547        3.547            0 0.729       0.729           0
## re75         -5.446        5.446            0 0.774       0.774           0
## 
## Summary of Balance for Weighted Data:
##              SMD TSMD Control TSMD Treated    KS TKS Control TKS Treated
## age            0            0            0 0.144       0.144           0
## education      0            0            0 0.085       0.085           0
## married        0            0            0 0.000       0.000           0
## raceblack      0            0            0 0.000       0.000           0
## racehispanic   0            0            0 0.000       0.000           0
## racewhite      0            0            0 0.000       0.000           0
## nodegree       0            0            0 0.000       0.000           0
## re74           0            0            0 0.593       0.593           0
## re75           0            0            0 0.579       0.579           0
## 
## Effective Sample Sizes:
##          Control Treated
## All      2490.       185
## Weighted  333.61     185

plot(lmw_att_sum)
\end{verbatim}
\doublespacing

\begin{figure}[H]
    \centering
    \includegraphics[width=4.5in]{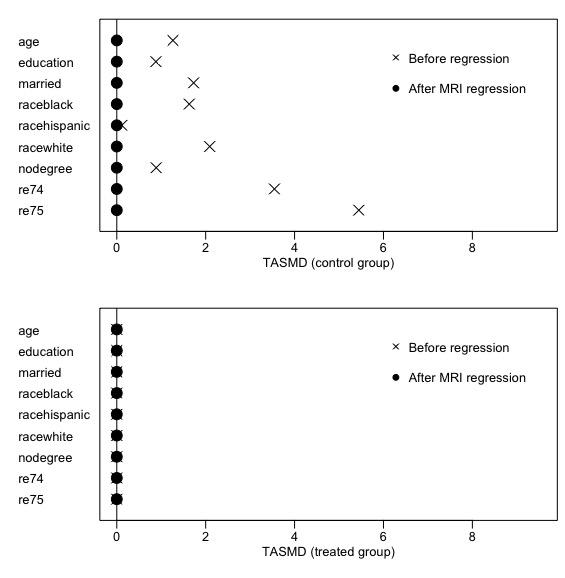}
\end{figure}

This time, the SMDs and TSMDs after weighting by the MRI weights are all exactly equal to \textcolor{black}{zero}, indicating that the \textcolor{black}{covariate means in each group are balanced exactly relative to the target (here, treated) population.} 
Some imbalances remain on features of the covariate distributions beyond the means, however, as indicated by the high KS statistics. The linearity of the model is not able to account for imbalances in the whole covariate distribution, despite perfectly balancing the means. Such imbalances can be particularly severe in the presence of negative weights.

We can use \texttt{plot()} with the argument \texttt{type = "weights"} to examine the distribution of the weights and, in particular, to examine the presence of negative weights.
\textcolor{black}{\texttt{lmw} automatically re-scales the weights in its output so that the sum of the weights within each group equals the corresponding group size. This implies that the re-scaled weights average to one within each group.}

\singlespacing
\begin{verbatim}
plot(lmw_att_out, type = "weights")
\end{verbatim}
\doublespacing

\begin{figure}[H]
    \centering
    \includegraphics[width=4.5in]{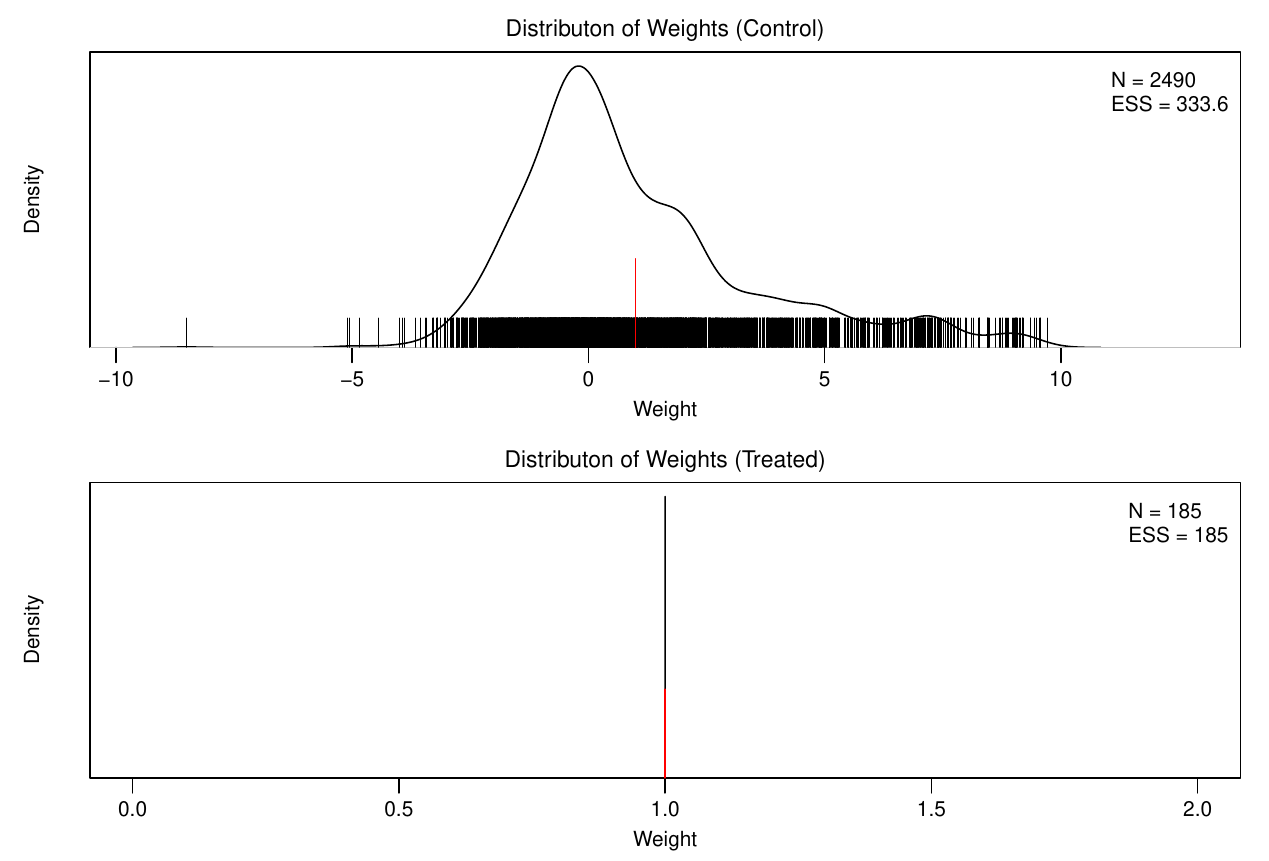}
\end{figure}

The upper and lower plots display the distributions of the weights for the control and treated groups, respectively (the treated group weights are constant and equal to 1). 
The distribution is depicted using a kernel density plot and a rug plot, with the vertical red line indicating the average of the weights (\textcolor{black}{which equals one after re-scaling}).
This plot indicates \textcolor{black}{considerable} extrapolation given that a large share of the weights distribution takes values less than 0 and with somewhat high magnitude. The high variability of the weights is reflected in the low ESS of 334 compared to the original sample size of 2490.

As before, one can request an extrapolation plot to further examine the extent of extrapolation using \texttt{plot()} with \texttt{type = "extrapolation"}:

\singlespacing
\begin{verbatim}
plot(lmw_att_out, type = "extrapolation",
     variables = ~age + married + re74)
\end{verbatim}
\doublespacing

\begin{figure}[H]
    \centering
    \includegraphics[width=4.5in]{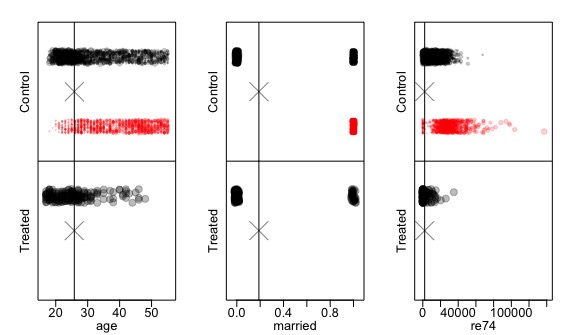}
\end{figure}

\textcolor{black}{The red circles in the plots again indicate negative weights and extrapolation of effect estimates.
This time, however, the weighted covariate means align exactly with the corresponding means in the target population, as indicated by the black vertical lines.
From these plots, observations with large, negative weights should be carefully examined for potential exclusion from the analysis, as they have substantially different covariate values from the observations in the counterpart treatment group. 
The findings from these plots also suggest the use of matching or weighting before regression to mitigate the non-negativity and extremity of the weights, as discussed in the next section.}
%

Finally, one can estimate the potential outcome means and the treatment effect using \texttt{lmw\_est()}, which fits the linear model specified in the \texttt{lmw} object. To do so, the outcome variable must be supplied to the \texttt{outcome} argument. The output of \texttt{lmw\_est()} contains the fitted model. To compute the effect estimate and its standard error, \texttt{summary()} must be called on the resulting output.

\singlespacing
\begin{Verbatim}[commandchars=\\\{\}]
lmw_att_est <- lmw_est(lmw_att_out, outcome = re78)
lmw_att_est
## An lmw_est object
##  - outcome: re78 
##  - standard errors: robust (HC3) 
##  - estimand: ATT 
##  - method: MRI 
## 
## Use summary() to examine estimates, standard errors, p-values,
## and confidence intervals.
summary(lmw_att_est)
## 
## Effect estimates:
##              Estimate Std. Error 95% CI L 95% CI U t value Pr(>|t|)
## {E[\m{\text{Y\textsubscript{1}-Y\textsubscript{0}}}|A=1]}    790.5      793.7   -765.7   2346.8   0.996    0.319
## 
## Residual standard error: 10060 on 2657 degrees of freedom
## 
## Potential outcome means:
##           Estimate Std. Error 95% CI L 95% CI U t value Pr(>|t|)    
## E[\m{\text{Y\textsubscript{0}}}|A=1]   5558.6      520.6   4537.8   6579.4   10.68   <2e-16 ***
## E[\m{\text{Y\textsubscript{1}}}|A=1]   6349.1      599.1   5174.5   7523.8   10.60   <2e-16 ***
## ---
## Signif. codes:  0 '***' 0.001 '**' 0.01 '*' 0.05 '.' 0.1 ' ' 1
\end{Verbatim}
\doublespacing

By default, \texttt{lmw\_est()} applies HC3 robust standard errors to the outcome model parameters; this \textcolor{black}{option} can be controlled using the \texttt{robust} argument. The \texttt{summary()} output \textcolor{black}{shows the estimated means of the potential outcomes under treatment and control as} \$6349 and \$5558, \textcolor{black}{respectively,} yielding an \textcolor{black}{estimated} ATT of \$791 (SE: \$794, 95\% CI: (-767, 2347), p = .319). 
When URI is used, the potential outcome means will not be displayed because they are not estimated by the outcome model directly; only the treatment effect is.

A common concern in weighting is that estimates may be dominated by a handful of observations, in particular for continuous outcomes such as earnings. The influence of individual observations can be assessed using the sample influence curve (SIC), which is a function of the implied regression weights, \textcolor{black}{leverages, and residuals based on} the fitted outcome model.\footnote{\textcolor{black}{More formally, the SIC for a treated unit $i$ under MRI is $\text{SIC}_i = (n_t-1)e_i w_i/(1-h_{ii,t})$, where $h_{ii,t}$ is the leverage of unit $i$, and $e_i$ is the corresponding residual from the fitted model in the treatment group.}}
The \texttt{influence()} function can be used on \texttt{lmw} or \texttt{lmw\_est} objects to produce the usual measures of influence (\textcolor{black}{leverages} and residuals) and SIC values. 
\textcolor{black}{A plot of the SIC values across all units can be produced using \texttt{plot()} with \texttt{type = "influence"}. The plot also highlights the indices of the most influential units.}

\singlespacing
\begin{verbatim}
plot(lmw_att_est, type = "influence")
\end{verbatim}
\doublespacing

\begin{figure}[H]
    \centering
    \includegraphics[width=4.5in]{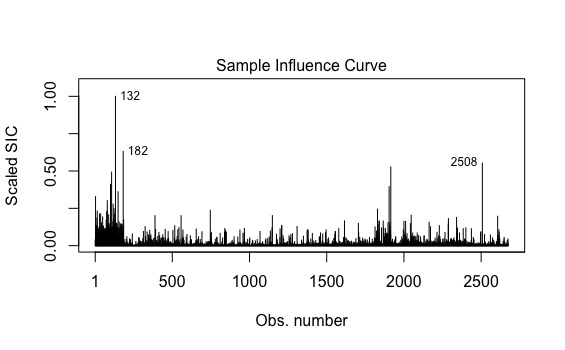}
\end{figure}

\subsection{MRI after matching for the ATT}
Performing regression after matching (e.g., propensity score matching) or weighting (e.g., inverse probability weighting) can reduce residual imbalances and extrapolation and add robustness to the effect estimate (i.e., robustness to misspecification of the outcome model). 
\texttt{lmw()} \textcolor{black}{has} a \texttt{base.weights} argument to incorporate balancing weights into the regression model for the outcome. When \texttt{base.weights} is supplied, the \texttt{dr.method} argument can be set to \texttt{"WLS"} to perform weighted least squares regression, or \texttt{"AIPW"} to perform augmented inverse probability weighting using these weights (\texttt{"WLS"} is the default). \texttt{lmw} also provides direct compatibility with the \texttt{MatchIt} and \texttt{WeightIt} packages for matching and weighting, respectively; the user can supply the output of a call to \texttt{MatchIt::matchit()} or \texttt{WeightIt::weightit()} to the \texttt{obj} argument of \texttt{lmw()}, which will extract the weights and \textcolor{black}{the} estimand.

Below, we use an initial matching step to attempt to balance the covariates prior to fitting \textcolor{black}{an} outcome model. We use \texttt{MatchIt::matchit()} to perform nearest neighbor propensity score matching.

\singlespacing
\begin{verbatim}
library(MatchIt)
matchit_out <- matchit(treat ~ age + education + married + race + 
                     nodegree + re74 + re75, data = lalonde,
                   estimand = "ATT")
matchit_out
## A matchit object
##  - method: 1:1 nearest neighbor matching without replacement
##  - distance: Propensity score
##              - estimated with logistic regression
##  - number of obs.: 2675 (original), 370 (matched)
##  - target estimand: ATT
##  - covariates: age, education, married, race, nodegree, re74, re75
\end{verbatim}
\doublespacing

Next, we supply the \texttt{matchit} output object to the \texttt{obj} argument of \texttt{lmw()} to incorporate the matching weights into the outcome model. This effectively fits a weighted least squares regression with the matching weights (in this case the matching weights are all either 0 or 1).

\singlespacing
\begin{verbatim}
lmw_match_out <- lmw(~ treat + age + education + married + race + 
                       nodegree + re74 + re75, obj = matchit_out,
                     method = "MRI")
lmw_match_out
## An lmw object
##  - treatment: treat (2 levels)
##  - method:  MRI (multi-regression imputation)
##  - number of obs.: 2675
##  - sampling weights: none
##  - base weights: matching weights from MatchIt
##  - doubly-robust method: weighted least squares (WLS)
##  - target estimand: ATT
##  - covariates: age, education, married, race, nodegree, re74, re75
\end{verbatim}
\doublespacing

When using \texttt{summary()} to assess balance, balance \textcolor{black}{tables are} presented for the unadjusted sample, the sample after matching, and the sample after matching and regression.

\singlespacing
\begin{verbatim}
summary(lmw_match_out)
## 
## Call:
## lmw(formula = ~treat + age + education + married + race + nodegree + 
##     re74 + re75, method = "MRI", obj = matchit_out)
## 
## Summary of Balance for Unweighted Data:
##                 SMD TSMD Control TSMD Treated    KS TKS Control TKS Treated
## age          -1.263        1.263            0 0.377       0.377           0
## education    -0.881        0.881            0 0.403       0.403           0
## married      -1.729        1.729            0 0.677       0.677           0
## raceblack     1.630       -1.630            0 0.593       0.593           0
## racehispanic  0.114       -0.114            0 0.027       0.027           0
## racewhite    -2.091        2.091            0 0.620       0.620           0
## nodegree      0.886       -0.886            0 0.403       0.403           0
## re74         -3.547        3.547            0 0.729       0.729           0
## re75         -5.446        5.446            0 0.774       0.774           0
## 
## Summary of Balance for Base Weighted Data:
##                 SMD TSMD Control TSMD Treated    KS TKS Control TKS Treated
## age          -0.652        0.652            0 0.178       0.178           0
## education    -0.016        0.016            0 0.092       0.092           0
## married      -0.690        0.690            0 0.270       0.270           0
## raceblack     0.238       -0.238            0 0.086       0.086           0
## racehispanic -0.023        0.023            0 0.005       0.005           0
## racewhite    -0.274        0.274            0 0.081       0.081           0
## nodegree      0.190       -0.190            0 0.086       0.086           0
## re74         -0.492        0.492            0 0.416       0.416           0
## re75         -0.519        0.519            0 0.297       0.297           0
## 
## Summary of Balance for Weighted Data:
##              SMD TSMD Control TSMD Treated    KS TKS Control TKS Treated
## age            0            0            0 0.147       0.147           0
## education      0            0            0 0.058       0.058           0
## married        0            0            0 0.000       0.000           0
## raceblack      0            0            0 0.000       0.000           0
## racehispanic   0            0            0 0.000       0.000           0
## racewhite      0            0            0 0.000       0.000           0
## nodegree       0            0            0 0.000       0.000           0
## re74           0            0            0 0.382       0.382           0
## re75           0            0            0 0.222       0.222           0
## 
## Effective Sample Sizes:
##               Control Treated
## All           2490.       185
## Base weighted  185.       185
## Weighted        77.93     185
\end{verbatim}
\doublespacing

Although balance after matching alone is somewhat poor, using MRI after matching improves balance \textcolor{black}{considerably}. \textcolor{black}{Also, MRI after matching typically has better balance than MRI} without matching \textcolor{black}{(see Section \ref{sec_mri_att})}. Matching can also decrease extrapolation, as the plots below indicate:

\singlespacing
\begin{verbatim}
plot(lmw_match_out, type = "extrapolation",
     variables = ~age + married + re74)
\end{verbatim}
\doublespacing

\begin{figure}[H]
    \centering
    \includegraphics[width=4.5in]{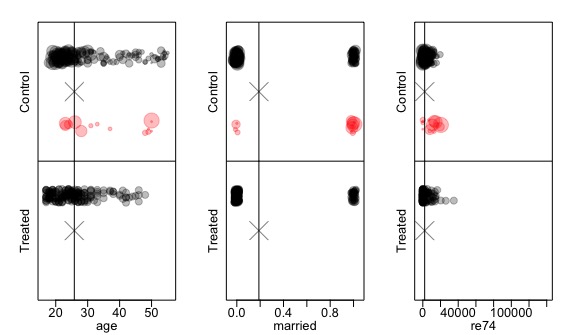}
\end{figure}

Finally, one can estimate the treatment effect, again using \texttt{lmw\_est()} and \texttt{summary()}. \textcolor{black}{Since} pair matching was used to find the matches, we use cluster-robust standard errors by setting  \texttt{cluster = \char`\~ subclass} in the call to \texttt{lmw\_est()}:

\singlespacing
\begin{Verbatim}[commandchars=\\\{\}]
lmw_match_est <- lmw_est(lmw_match_out, outcome = re78, cluster = ~subclass)
lmw_match_est
## An lmw_est object
##  - outcome: re78 
##  - standard errors: cluster robust (HC1) 
##  - estimand: ATT 
##  - method: MRI 
## 
## Use summary() to examine estimates, standard errors, p-values, 
## and confidence intervals.
summary(lmw_match_est)
## 
## Effect estimates:
##              Estimate Std. Error 95% CI L 95% CI U t value Pr(>|t|)  
## E[\m{\text{Y}\textsubscript{1}\text{-}\text{Y}\textsubscript{0}}|A=1]   1904.5      872.3    189.0   3620.1   2.183   0.0297 *
## ---
## Signif. codes:  0 '***' 0.001 '**' 0.01 '*' 0.05 '.' 0.1 ' ' 1
## 
## Residual standard error: 6922 on 352 degrees of freedom
## 
## Potential outcome means:
##           Estimate Std. Error 95% CI L 95% CI U t value Pr(>|t|)    
## E[\m{\text{Y}\textsubscript{0}}|A=1]   4444.6      634.0   3197.8   5691.4   7.011 1.22e-11 ***
## E[\m{\text{Y}\textsubscript{1}}|A=1]   6349.1      577.2   5213.9   7484.4  10.999  < 2e-16 ***
## ---
## Signif. codes:  0 '***' 0.001 '**' 0.01 '*' 0.05 '.' 0.1 ' ' 1
\end{Verbatim}
\doublespacing

The \texttt{summary()} output \textcolor{black}{shows that the estimated potential outcomes means under treatment and control are \$6349 and \$4445, respectively}, yielding an \textcolor{black}{estimated} ATT of \$1905 (SE: \$872, 95\% CI: (189, 3620), p = .030).

\subsection{Multi-valued treatments}
\textcolor{black}{Both URI and MRI extend naturally to settings with multi-valued treatments. In this section, we demonstrate the use of MRI to estimate the average effect of the treatment (treatment 1) with respect to the two artificially constructed controls (treatment 2 and treatment 3; see Section \ref{sec_data}).}
The syntax for multi-valued treatments is essentially the same as with binary treatments except that the \texttt{focal} argument needs to be supplied when \texttt{estimand = "ATT"} to identify which treatment level is to be considered the “treated” or “focal” level. (When URI is used, an additional \texttt{contrast} argument is required to identify a pair of groups to be contrasted, since each contrast will receive its own set of weights, which will be computed one at a time by \texttt{lmw()}.)

\singlespacing
\begin{verbatim}
lmw_multi_out <- lmw(~ treat_multi + age + education + married + race + 
                     nodegree + re74 + re75, data = lalonde,
                   estimand = "ATT", method = "MRI",
                   treat = "treat_multi", focal = "1")
lmw_multi_out
## An lmw object
##  - treatment: treat_multi (3 levels)
##  - method:  MRI (multi-regression imputation)
##  - number of obs.: 2675
##  - sampling weights: none
##  - base weights: none
##  - target estimand: ATT (focal = "1")
##  - covariates: age, education, married, race, nodegree, re74, re75
\end{verbatim}
\doublespacing

Using \texttt{summary()} on the resulting object produces a balance table for each treatment level \textcolor{black}{relative to} the target (in this case, treatment level 1). The \texttt{contrast} argument can be used to specify for which pair of levels balance statistics should be displayed (when using URI, only the two groups involved in the contrast supplied to \texttt{lmw()} will be compared).

\singlespacing
\begin{verbatim}
summary(lmw_multi_out)
## 
## Call:
## lmw(formula = ~treat_multi + age + education + married + race + 
##     nodegree + re74 + re75, data = lalonde, estimand = "ATT", 
##     method = "MRI", treat = "treat_multi", focal = "1")
## 
## Summary of Balance for Unweighted Data:
##              TSMD 1 TSMD 2 TSMD 3 TKS 1 TKS 2 TKS 3
## age               0  1.791  0.963     0 0.538 0.312
## education         0 -0.210  1.499     0 0.164 0.573
## married           0  1.679  1.757     0 0.658 0.688
## raceblack         0 -0.912 -2.037     0 0.332 0.741
## racehispanic      0 -0.101 -0.121     0 0.024 0.029
## racewhite         0  1.200  2.596     0 0.356 0.769
## nodegree          0 -0.225 -1.261     0 0.102 0.573
## re74              0  2.350  4.226     0 0.634 0.791
## re75              0  3.297  6.664     0 0.630 0.857
## 
## Summary of Balance for Weighted Data:
##              TSMD 1 TSMD 2 TSMD 3 TKS 1 TKS 2 TKS 3
## age               0      0      0     0 0.158 0.118
## education         0      0      0     0 0.028 0.086
## married           0      0      0     0 0.000 0.000
## raceblack         0      0      0     0 0.000 0.000
## racehispanic      0      0      0     0 0.000 0.000
## racewhite         0      0      0     0 0.000 0.000
## nodegree          0      0      0     0 0.000 0.000
## re74              0      0      0     0 0.484 0.677
## re75              0      0      0     0 0.439 0.675
## 
## Effective Sample Sizes:
##            1      2       3
## All      185 901.   1589.  
## Weighted 185 153.16  109.55
\end{verbatim}
\doublespacing

\textcolor{black}{With the MRI, the covariate} means in each treatment group are \textcolor{black}{exactly balanced relative to the target}, but some imbalances remain on the KS statistics. As with binary treatments, \texttt{plot()} can be used on the \texttt{summary()} output to \textcolor{black}{generate balance plots, or} on the \texttt{lmw()} output to \textcolor{black}{generate plots of} the distribution of weights and the degree of extrapolation.

Estimating the treatment effect follows the same procedure as with binary treatments: \textcolor{black}{we can} use \texttt{lmw\_est()} to fit the outcome model and use \texttt{summary()} to extract the treatment effect estimates and their standard errors.

\singlespacing
\begin{Verbatim}[commandchars=\\\{\}]
lmw_multi_est <- lmw_est(lmw_multi_out, outcome = re78)
lmw_multi_est
## An lmw_est object
##  - outcome: re78 
##  - standard errors: robust (HC3) 
##  - estimand: ATT 
##  - method: MRI 
## 
## Use summary() to examine estimates, standard errors, p-values, 
## and confidence intervals.
summary(lmw_multi_est)
## 
## Effect estimates:
##              Estimate Std. Error 95% CI L 95% CI U t value Pr(>|t|)
## E[\m{\text{Y}\textsubscript{1}\text{-}\text{Y}\textsubscript{2}}|A=1]   1418.5      866.1   -279.7   3116.7   1.638    0.102
## E[\m{\text{Y}\textsubscript{1}\text{-}\text{Y}\textsubscript{3}}|A=1]    757.6     1114.6  -1427.9   2943.1   0.680    0.497
## E[\m{\text{Y}\textsubscript{2}\text{-}\text{Y}\textsubscript{3}}|A=1]   -660.9     1129.0  -2874.6   1552.9  -0.585    0.558
## 
## Residual standard error: 10040 on 2648 degrees of freedom
## 
## Potential outcome means:
##           Estimate Std. Error 95% CI L 95% CI U t value Pr(>|t|)    
## E[\m{\text{Y}\textsubscript{1}}|A=1]   6349.1      599.1   5174.5   7523.8  10.598  < 2e-16 ***
## E[\m{\text{Y}\textsubscript{2}}|A=1]   4930.6      625.5   3704.2   6157.1   7.883 4.62e-15 ***
## E[\m{\text{Y}\textsubscript{3}}|A=1]   5591.5      939.9   3748.6   7434.5   5.949 3.05e-09 ***
## ---
## Signif. codes:  0 '***' 0.001 '**' 0.01 '*' 0.05 '.' 0.1 ' ' 1
\end{Verbatim}
\doublespacing

The \textcolor{black}{estimated} ATT \textcolor{black}{of treatment 1 relative to 2, and 1 relative to 3,} are \$1419 (SE: \$866, 95\% CI: (-280, 3117), p = .102) and \$758 (SE: \$1115, 95\% CI: (-1428, 2943), p = .497), \textcolor{black}{respectively}.

\subsection{\textcolor{black}{Instrumental variables}}
Using \texttt{lmw} with instrumental variables is straightforward; the user must specify the “second-stage” (i.e., outcome) model in the main model formula and must supply the instrument separately. The \texttt{lmw\_iv()} function is used to compute the weights implied by the 2SLS model. Both URI and MRI are possible with 2SLS, although URI is far more common. When MRI is used or treatment-covariate interactions are included in the outcome model, interactions between the covariates and the instrument will be included as additional instruments.

\textcolor{black}{In this section, we demonstrate the use of URI}, using the simulated variable \texttt{Ins} as an instrument and \textcolor{black}{adjusting for the covariates} \texttt{age}, \texttt{education}, \texttt{race}, and \texttt{re74} in the 2SLS models. 

\singlespacing
\begin{verbatim}
lmw_iv_out <- lmw_iv(~treat + age + education + race + re74,
                     data = lalonde, estimand = "ATT", method = "URI",
                     treat = "treat", iv = ~Ins)
lmw_iv_out
## An lmw_iv object
##  - treatment: treat (2 levels)
##  - instrument: Ins
##  - method: URI (uni-regression imputation)
##  - number of obs.: 2675
##  - sampling weights: none
##  - base weights: none
##  - target estimand: ATT
##  - covariates: age, education, race, re74
\end{verbatim}
\doublespacing

Specifying the estimand does not affect estimation of the weights with URI when no treatment-covariate interactions are included in the outcome model, but it can be useful to determine how far the target population implied by the weights is from the desired target population.

As before, one can use \texttt{summary()} to assess balance. Variables that were not included in the model can be added using \texttt{addlvariables}.

\singlespacing
\begin{verbatim}
summary(lmw_iv_out, addlvariables = ~married + nodegree + re75)
## 
## Call:
## lmw_iv(formula = ~treat + age + education + race + re74, data = lalonde, 
##     estimand = "ATT", method = "URI", treat = "treat", iv = ~Ins)
## 
## Summary of Balance for Unweighted Data:
##                 SMD TSMD Control TSMD Treated    KS TKS Control TKS Treated
## age          -1.263        1.263            0 0.377       0.377           0
## education    -0.881        0.881            0 0.403       0.403           0
## raceblack     1.630       -1.630            0 0.593       0.593           0
## racehispanic  0.114       -0.114            0 0.027       0.027           0
## racewhite    -2.091        2.091            0 0.620       0.620           0
## re74         -3.547        3.547            0 0.729       0.729           0
## married      -1.729        1.729            0 0.677       0.677           0
## nodegree      0.886       -0.886            0 0.403       0.403           0
## re75         -5.446        5.446            0 0.774       0.774           0
## 
## Summary of Balance for Weighted Data:
##                 SMD TSMD Control TSMD Treated    KS TKS Control TKS Treated
## age           0.000        0.539        0.539 0.324       0.432       0.198
## education     0.000       -0.218       -0.218 0.580       0.534       0.106
## raceblack     0.000       -0.557       -0.557 0.000       0.203       0.203
## racehispanic  0.000        0.390        0.390 0.000       0.092       0.092
## racewhite     0.000        0.373        0.373 0.000       0.110       0.110
## re74          0.000       -0.219       -0.219 0.704       0.725       0.103
## married      -1.255        1.388        0.134 0.491       0.544       0.052
## nodegree      1.275       -1.175        0.100 0.580       0.534       0.045
## re75          0.260       -0.322       -0.062 0.535       0.532       0.059
## 
## Effective Sample Sizes:
##          Control Treated
## All      2490.     185. 
## Weighted    2.25    59.6
\end{verbatim}
\doublespacing

Extrapolation can be a problem with 2SLS models due to large, negative weights; \textcolor{black}{as in standard URI and MRI, such extrapolation can be assessed graphically using an extrapolation plot of the weights.}

\singlespacing
\begin{verbatim}
plot(lmw_iv_out, type = "extrapolation",
     variables = ~age + married + re74)
\end{verbatim}
\doublespacing

\textcolor{black}{If there are observations with large, negative weights, it is advisable to closely scrutinize them and consider excluding them from the analyses.
In essence, these are observations with substantially different covariate values from those in the alternate treatment group.
As an alternative strategy, one may consider doing matching or weighting adjustments before regression, thereby constraining the weights to non-negative values, zeroing out such disparate observations, and promoting the estimator to be sample-bounded.}

Finally, one can estimate the treatment effect using \texttt{lmw\_est()}, which produces an estimate based on the second-stage regression model on the treatment and covariates after adjustment by the instrument. As with other regression models, robust standard errors are produced by default.

\singlespacing
\begin{Verbatim}[commandchars=\\\{\}]
lmw_iv_est <- lmw_est(lmw_iv_out, outcome = re78)
lmw_iv_est
## An lmw_est_iv object
##  - outcome: re78 
##  - standard errors: robust (HC3) 
##  - estimand: ATT 
##  - method: URI 
## 
## Use summary() to examine estimates, standard errors, p-values, 
## and confidence intervals.
summary(lmw_iv_est)
## 
## Effect estimates:
##              Estimate Std. Error 95% CI L 95% CI U t value Pr(>|t|)
## E[\m{\text{Y}\textsubscript{1}\text{-}\text{Y}\textsubscript{0}}|A=1]     7802       7160    -6238    21841    1.09    0.276
## 
## Residual standard error: 11030 on 2668 degrees of freedom
\end{Verbatim}
\doublespacing

The \textcolor{black}{estimated} treatment effect is \$7802 (SE: \$7160, 95\% CI: (-6238, 21841), p = .276). Note this estimate should not be taken seriously as the instrument here was synthetic and used solely for illustration purposes.

\textcolor{black}{\section{Remarks}}
\label{sec_packages}

\textcolor{black}{The linear regression model is extensively used in observational studies of causal effects.
However, in its traditional implementation via the least squares method, for instance in the \texttt{lm} package in \texttt{R}, it is not clear how the model uses each of the individual observations in the data to produce an effect estimate, and how well its adjustments approximate features of a hypothetical randomized experiment.
In this work, we have illustrated an alternative yet equivalent implementation of the linear regression model in the new \texttt{lmw} package in \texttt{R}, which enables the computation of the exact, implied weights of the model.
Crucially, the weights are computed in the design stage of the study, specifically without using outcome information.
The weights provide insight into the hypothetical experiment underpinning linear regression in terms of covariate balance, study representativeness, interpolated estimation, and unweighted analyses \citep{chattopadhyay2024causation}.} 

\textcolor{black}{The \texttt{lmw} package has three primary functions: weight computation, weight assessment, and effect estimation. 
To the best of our knowledge, this is the first package to offer these functionalities in the context of outcome modeling.
Other packages such as \texttt{PSweight} and \texttt{WeightIt}  compute weights in the context of treatment or propensity score weighting. 
\texttt{sbw} and \texttt{optweight} estimate stable balancing weights \citep{zubizarreta2015stable}, of which linear model weights can be seen as a special case (i.e., with exact balance on the means, specific covariate profiles, and allowance for negative weights). 
\texttt{cobalt} provides an interface for assessing covariate balance after weighting, but \texttt{lmw} provides specific tools for assessing balance, representativeness, extrapolation, and influence using the framework of linear model weights. 
For estimating treatment effects, \texttt{PSweight} supports g-computation and AIPW, and \texttt{ivmodel} and \texttt{ivreg} support 2SLS.
\texttt{lmw} also provides functionalities to estimate treatment effects using these methods and is benchmarked against these packages to ensure results are consistent among them when applicable.}

\pagebreak
\bibliographystyle{asa}
\bibliography{bib}

\begin{thebibliography}{32}
\newcommand{\enquote}[1]{``#1''}
\expandafter\ifx\csname natexlab\endcsname\relax\def\natexlab#1{#1}\fi

\bibitem[{Abadie et~al.(2015)Abadie, Diamond, and
  Hainmueller}]{abadie2015comparative}
Abadie, A., Diamond, A., and Hainmueller, J. (2015), \enquote{Comparative
  politics and the synthetic control method,} \textit{American Journal of
  Political Science}, 59, 495--510.

\bibitem[{Angrist and Pischke(2008)}]{angrist2008mostly}
Angrist, J.~D. and Pischke, J.-S. (2008), \textit{Mostly harmless econometrics:
  An empiricist's companion}, Princeton university press.

\bibitem[{Aronow and Samii(2016)}]{aronow2016does}
Aronow, P.~M. and Samii, C. (2016), \enquote{Does regression produce
  representative estimates of causal effects?} \textit{American Journal of
  Political Science}, 60, 250--267.

\bibitem[{Austin and Stuart(2015)}]{austin2015moving}
Austin, P.~C. and Stuart, E.~A. (2015), \enquote{Moving towards best practice
  when using inverse probability of treatment weighting (IPTW) using the
  propensity score to estimate causal treatment effects in observational
  studies,} \textit{Statistics in Medicine}, 34, 3661--3679.

\bibitem[{Baiocchi et~al.(2014)Baiocchi, Cheng, and
  Small}]{baiocchi2014instrumental}
Baiocchi, M., Cheng, J., and Small, D.~S. (2014), \enquote{Instrumental
  variable methods for causal inference,} \textit{Statistics in Medicine}, 33,
  2297--2340.

\bibitem[{Ben-Michael et~al.(2021)Ben-Michael, Feller, and
  Rothstein}]{ben2021augmented}
Ben-Michael, E., Feller, A., and Rothstein, J. (2021), \enquote{The augmented
  synthetic control method,} \textit{Journal of the American Statistical
  Association}, 116, 1789--1803.

\bibitem[{Chattopadhyay(2022)}]{chattopadhyay2022new}
Chattopadhyay, A. (2022), \enquote{New Methods for Causal Inference in
  Randomized Experiments and Observational Studies,} Ph.D. thesis, Harvard
  University.

\bibitem[{Chattopadhyay et~al.(2023)Chattopadhyay, Cohn, and
  Zubizarreta}]{chattopadhyay2023one}
Chattopadhyay, A., Cohn, E.~R., and Zubizarreta, J.~R. (2023),
  \enquote{One-step weighting to generalize and transport treatment effect
  estimates to a target population,} \textit{The American Statistician}, 1--10.

\bibitem[{Chattopadhyay et~al.(2020)Chattopadhyay, Hase, and
  Zubizarreta}]{chattopadhyay2020balancing}
Chattopadhyay, A., Hase, C.~H., and Zubizarreta, J.~R. (2020),
  \enquote{Balancing vs modeling approaches to weighting in practice,}
  \textit{Statistics in Medicine}, 39, 3227--3254.

\bibitem[{Chattopadhyay and Zubizarreta(2023)}]{chattopadhyay2023implied}
Chattopadhyay, A. and Zubizarreta, J.~R. (2023), \enquote{On the implied
  weights of linear regression for causal inference,} \textit{Biometrika}, 110,
  615--629.

\bibitem[{Chattopadhyay and Zubizarreta(2024)}]{chattopadhyay2024causation}
--- (2024), \enquote{{Causation}, {Comparison}, and {Regression},}
  \textit{Harvard Data Science Review}, 6,
  https://hdsr.mitpress.mit.edu/pub/1ybwbmlw.

\bibitem[{Cochran(1965)}]{cochran1965planning}
Cochran, W.~G. (1965), \enquote{The planning of observational studies of human
  populations,} \textit{Journal of the Royal Statistical Society}, 128,
  234--266.

\bibitem[{Cohn and Zubizarreta(2022)}]{cohn2021profile}
Cohn, E.~R. and Zubizarreta, J.~R. (2022), \enquote{Profile Matching for the
  Generalization and Personalization of Causal Inferences,} \textit{arXiv
  prep"rint arXiv:2105.10060v3}.

\bibitem[{Dehejia and Wahba(1999)}]{dehejia1999causal}
Dehejia, R. and Wahba, S. (1999), \enquote{Causal effects in nonexperimental
  studies: Reevaluating the evaluation of training programs,} \textit{Journal
  of the American Statistical Association}, 94, 1053--1062.

\bibitem[{Dorn(1953)}]{dorn1953philosophy}
Dorn, H.~F. (1953), \enquote{Philosophy of inferences from retrospective
  studies,} \textit{American Journal of Public Health and the Nations Health},
  43, 677--683.

\bibitem[{Gelman and Imbens(2018)}]{gelman2018high}
Gelman, A. and Imbens, G. (2018), \enquote{Why high-order polynomials should
  not be used in regression discontinuity designs,} \textit{Journal of Business
  \& Economic Statistics}, 1--10.

\bibitem[{Greifer and Stuart(2021)}]{greifer2021choosing}
Greifer, N. and Stuart, E.~A. (2021), \enquote{Choosing the estimand when
  matching or weighting in observational studies,} \textit{arXiv preprint
  arXiv:2106.10577}.

\bibitem[{Hern{\'a}n and Robins(2016)}]{hernan2016using}
Hern{\'a}n, M.~A. and Robins, J.~M. (2016), \enquote{Using big data to emulate
  a target trial when a randomized trial is not available,} \textit{American
  Journal of Epidemiology}, 183, 758--764.

\bibitem[{Hern\'an and Robins(2020)}]{hernan2020causal}
Hern\'an, M.~A. and Robins, J.~M. (2020), \textit{Causal inference: What if},
  CRC Boca Raton.

\bibitem[{Imbens(2015)}]{imbens2015matching}
Imbens, G.~W. (2015), \enquote{Matching methods in practice: three examples,}
  \textit{Journal of Human Resources}, 50, 373--419.

\bibitem[{Imbens and Rubin(2015)}]{imbens2015causal}
Imbens, G.~W. and Rubin, D.~B. (2015), \textit{Causal inference in statistics,
  social, and biomedical sciences}, Cambridge University Press.

\bibitem[{LaLonde(1986)}]{lalonde1986evaluating}
LaLonde, R.~J. (1986), \enquote{Evaluating the econometric evaluations of
  training programs with experimental data,} \textit{The American Economic
  Review}, 604--620.

\bibitem[{Neyman(1923,1990)}]{neyman1923application}
Neyman, J. (1923,1990), \enquote{On the application of probability theory to
  agricultural experiments,} \textit{Statistical Science}, 5, 463--480.

\bibitem[{Robins et~al.(2007)Robins, Sued, Lei-Gomez, and
  Rotnitzky}]{robins2007comment}
Robins, J., Sued, M., Lei-Gomez, Q., and Rotnitzky, A. (2007),
  \enquote{Comment: performance of double-robust estimators when" inverse
  probability" weights are highly variable,} \textit{Statistical Science}, 22,
  544--559.

\bibitem[{Rosenbaum(2010)}]{rosenbaum2010design1}
Rosenbaum, P.~R. (2010), \textit{Design of observational studies}, Springer.

\bibitem[{Rosenbaum and Rubin(1983)}]{rosenbaum1983central}
Rosenbaum, P.~R. and Rubin, D.~B. (1983), \enquote{The central role of the
  propensity score in observational studies for causal effects,}
  \textit{Biometrika}, 70, 41--55.

\bibitem[{Rosenbaum and Rubin(1985)}]{rosenbaum1985constructing}
--- (1985), \enquote{Constructing a control group using multivariate matched
  sampling methods that incorporate the propensity score,} \textit{The American
  Statistician}, 39, 33--38.

\bibitem[{Rubin(1974)}]{rubin1974estimating}
Rubin, D.~B. (1974), \enquote{Estimating causal effects of treatments in
  randomized and nonrandomized studies.} \textit{Journal of Educational
  Psychology}, 66, 688.

\bibitem[{Rubin(2007)}]{rubin2007design}
--- (2007), \enquote{The design versus the analysis of observational studies
  for causal effects: parallels with the design of randomized trials,}
  \textit{Statistics in Medicine}, 26, 20--36.

\bibitem[{Schafer and Kang(2008)}]{schaferAverageCausalEffects2008}
Schafer, J.~L. and Kang, J. (2008), \enquote{Average causal effects from
  nonrandomized studies: {A} practical guide and simulated example,}
  \textit{Psychological Methods}, 13, 279--313.

\bibitem[{Terza et~al.(2008)Terza, Basu, and Rathouz}]{terza2008two}
Terza, J.~V., Basu, A., and Rathouz, P.~J. (2008), \enquote{Two-stage residual
  inclusion estimation: addressing endogeneity in health econometric modeling,}
  \textit{Journal of health economics}, 27, 531--543.

\bibitem[{Zubizarreta(2015)}]{zubizarreta2015stable}
Zubizarreta, J.~R. (2015), \enquote{Stable weights that balance covariates for
  estimation with incomplete outcome data,} \textit{Journal of the American
  Statistical Association}, 110, 910--922.

\end{thebibliography}


\end{document}